\documentclass[aps,showpacs]{revtex4}
\usepackage{graphicx}
\usepackage{amssymb}
\usepackage{amsmath}
\newcommand{\be}{\begin{equation}}
\newcommand{\ee}{\end{equation}}
\newcommand{\ben}{\begin{eqnarray}}
\newcommand{\een}{\end{eqnarray}}
\newcommand{\bes}{\begin{subequations}}
\newcommand{\ees}{\end{subequations}}
\newcommand{\bb}{\bibitem}

\begin{document}
\title{Duality linking standard and tachyon scalar field cosmologies}
\author{P.P. Avelino}
\affiliation{Centro de F\'{\i}sica do Porto, Rua do Campo Alegre 687, 4169-007 Porto, Portugal}
\affiliation{Departamento de F\'{\i}sica da Faculdade de Ci\^encias
da Universidade do Porto, Rua do Campo Alegre 687, 4169-007 Porto, Portugal}
\author{D. Bazeia}
\affiliation{Departamento de F\'{\i}sica, Universidade Federal da Para\'{\i}ba
58051-970 Jo\~ao Pessoa, Para\'{\i}ba, Brasil}
\author{L. Losano}
\affiliation{Departamento de F\'{\i}sica, Universidade Federal da Para\'{\i}ba
58051-970 Jo\~ao Pessoa, Para\'{\i}ba, Brasil}
\author{J.C.R.E. Oliveira}
\affiliation{Centro de F\'{\i}sica do Porto, Rua do Campo Alegre 687, 4169-007 Porto, Portugal}
\affiliation{Departamento de Engenharia F\'{\i}sica da Faculdade de Engenharia
da Universidade do Porto, Rua Dr. Roberto Frias, s/n, 4200-465 Porto, Portugal}
\author{A.B. Pavan}
\affiliation{Instituto de F\'\i sica, Universidade de S\~ao Paulo, Caixa Postal 66318, 05315-970 S\~ao Paulo SP, Brasil}
\date{\today}

\begin{abstract}
In this work we investigate the duality linking standard and tachyon scalar field cosmologies.  
We determine the transformation between standard and tachyon scalar fields and between 
their associated potentials, corresponding to the same background evolution. We show that, in 
general, the duality is broken at a perturbative level, when deviations from a homogeneous and 
isotropic background are taken into account. However, we find that for slow-rolling fields the duality 
is still preserved at a linear level. We illustrate our results with specific examples of cosmological 
relevance, where the correspondence between scalar and tachyon scalar field models 
can be calculated explicitly.
\end{abstract}


\maketitle


\section{Introduction}

The study of cosmic duality has shown to be very useful in particular in the understanding of the correspondence between different families of inflationary and dark energy models. Also, in field theory and string theory dualities have been used to obtain quantitative predictions in various limits \cite{lid}, in some cases with interesting implications to cosmology \cite{ven}.  In a Friedmann-Robertson-Walker Universe (FRW) the equations of motion for the evolution of the total density field are invariant under the duality transformation linking a standard expanding cosmology dominated by quintessence and a contracting cosmology with phantom behavior \cite{chi}. This duality has been generalized in order to include more complex dark energy models, in particular those where an interaction between dark matter and dark energy fluids is present. In all these models the duality may occur at the background level but is generally broken at a perturbative level. 

Cosmological scalar fields are expected to be relevant both at early and late times both in the context of primordial inflation and dark energy, respectively. In \cite{pad} Padmanabham noted a possible correspondence between a minimally coupled quintessence scalar field and a tachyon field described by a specific Lagrangian, suggested by Sen \cite{sen1,sen2}, in the context of string theory. Such correspondence was also explored by Gorini et al. in \cite{go}, which explicitly demonstrated that distinct scalar field and tachyon models may indeed give rise to the very same cosmological evolution, provided some general assumptions are satisfied \cite{pad,go,pad2}. In the present work we return to this issue, focusing on a large class of models for which the first-order formalism described in  \cite{bglm} can be applied. This framework has been very useful to investigate exact analytical solutions in the context of supergravity \cite{sugra}, branes \cite{B1,B2} and more recently in cosmology \cite{bglm,BC1}.

The outline of this paper is as follows. In Sec. II we consider FRW models with a generic scalar field and show that there is always an infinite set of scalar field models consistent with the same background evolution but, in general, with very different sound speeds. In Sec. III we explore, at the background level, the duality between standard quintessence and tachyon scalar field cosmologies.  In Sec. IV we investigate the correspondence between standard and tachyon scalar field cosmologies in an inflationary phase in the early universe, considering not only the background evolution but also the spectra of scalar and 
tensor perturbations generated during inflation. In the end of the section we give two specific examples where the 
correspondence between scalar and tachyon scalar field models can be calculated explicitly. Throughout this work we 
use units in which ${4\pi G}=1$.

\section{FRW models with a generic scalar field}

In this paper we shall consider models with a real scalar field, $\chi$, minimally coupled to gravity with the action
\be\label{model}
S=\int\,d^4x\;{\sqrt{-g}\;\left(-\frac14\,R+{\mathcal L(\chi,X)}\right)}\,,
\ee
where $X=\chi_{,\mu}  \chi^{,\mu}/2$. The energy-momentum tensor of the real scalar field can be written 
in a perfect fluid form
\begin{equation}\label{eq:fluid}
T^{\mu\nu}= (\rho+ p) u^\mu u^\nu - p g^{\mu\nu} \,,
\end{equation}
by means of the following identifications
\begin{equation}\label{eq:new_identifications}
u_\mu = \frac{\chi_{, \mu}}{\sqrt{2X}} \,,  \quad \rho = 2 X {\mathcal L}_{,X} - {\mathcal L} \, ,\quad p =  {\mathcal L}(X,\chi)\,.
\end{equation}
In Eq.~(\ref {eq:fluid}), $u^\mu$ is the 4-velocity field describing the motion of the fluid (for timelike $\chi_{, \mu}$), while $\rho$ and $p$ are its proper energy density and pressure, respectively. The equation of state parameter, $w$ is 
\begin{equation} 
\label{eq:w}
w \equiv \frac{p}{\rho} = \frac{\mathcal L}{2X {\mathcal L}_{,X}  - {\mathcal L}}\,, 
\end{equation} 
and, if ${\mathcal L}_{,X}\ne 0$, the sound speed squared is given by
\begin{equation}
\label{eq:cs2}
c_s^2 \equiv \frac{p_{,X}}{\rho_{,X}}=\frac{{\mathcal L}_{,X}}{{\mathcal L}_{,X}+2X{\mathcal L}_{,XX}}\,.
\end{equation}
If $X$ is a small quantity, compared to the energy density associated with the scalar field potential, then a generic Lagrangian 
is expected to admit an expansion of the form \cite{pdm}
\begin{equation}\label{lagran}
{\mathcal L}=-V(\chi)+f(\chi)X+g(\chi) X^2+...\,,
\end{equation}
where $f$ and $g$ are functions of $\chi$. Consequently,
\begin{equation}\label{cs}
c_s^2=1-4\frac{g}{f}X+...\,,
\end{equation}
where we are implicitly assuming that $f \neq 0$ for all relevant values of the scalar field $\chi$. Hence, $c_s^2 \sim 1$ as 
long as second and higher order kinetic terms can be neglected in Eq. (\ref {lagran}).

Consider a flat Friedmann-Robertson-Walker background with line element
\be
ds^2=dt^2 - a^2(t)\left(dx^2+dy^2 +dz^2\right) \,,
\ee
where $t$ is the physical time and $x$, $y$ and $z$ are comoving spatial coordinates.
Einstein's equations then imply 
\bes
\ben
\label{eq:ein1}
H^2&=&\frac23\,\rho\,,
\\
\label{eq:ein2}
{\dot H}&=&-(\rho+p)\,,
\een
\ees
where $H={\dot a}/a$ is the Hubble parameter and a dot represents a derivative with respect to physical time.
Energy-momentum conservation leads to
\be
\label{eq:ein3}
{\dot \rho} = -3H(\rho+p) \,.
\ee
This is not an independent equation since it can be obtained using Eqs.~(\ref {eq:ein1}) and (\ref {eq:ein2}). 
Given initial conditions for $\rho$ the dynamics of the universe is completely determined by $w(a)$. 
However, this is insufficient to determine ${\mathcal L(\chi,X)}$ since there is always an infinite set of scalar field models 
consistent with a given background evolution. This may be accomplished for example by adjusting a potential for the 
scalar field, as long as the model allows for the given $w(a)$.

On the other hand, the background dynamics fully determines the equation of state parameter
\be
w=-1 +\frac23 \frac{\dot H}{H} \,.
\ee
The same applies to the sound speed, but only if the pressure is a function of the density alone ($p = p(\rho)$). In that case
\be
c_s^2=-1 +\frac13 \frac{\ddot H}{{\dot H}H} \,.
\ee
For slow-rolling fields this would lead to $c_s^2 \sim -1$ which would generate instabilities in the model. However, in general, the background dynamics does not determine the sound speed as will be shown in the following section.

\section{Standard and tachyon scalar field cosmologies}

In this paper we shall investigate the duality between two families of scalar field models described by the Lagrangians 
\bes
\ben
\label{td}
{\mathcal L}=\frac12 \phi_{,\mu} \phi^{,\mu}-V(\phi)\,,
\\
{\mathcal {L}}=-U(\psi)\sqrt{1-\psi_{,\mu} \psi^{,\mu}}\,,
\een
\ees
where $V$ and $U$ are the potentials for the standard and tachyonic real scalar fields, $\phi$ and $\psi$ respectively. The equation of state parameters for the standard and tachyon fields are 
\bes
\ben
w_\phi=\frac{{\dot \phi}^2/2 - V}{{\dot \phi}^2/2 +V}\,,
\\
w_\psi=-1+{\dot \psi}^2\,,
\een
\ees
assuming they are homogeneous. On the other hand, the sound speed squared is $c_s^2=1$ for the standard scalar field and $c_s^2=-w$ in the case of the tachyon field. This makes the tachyon field an interesting unified dark energy candidate since 
$c_s^2 \to 0$ when $w \to 0$ allowing structures to grow in that limit. In fact, if $U$ is a constant then the tachyon model reduces 
to the standard Chaplygin gas model \cite{kmp}. Note that ${\dot \psi}^2 \le 1$ and consequently $-1 \le w_\psi \le 0$, while 
 $-1 \le w_\phi \le 1$. Hence, the duality between standard and tachyon scalar field cosmologies can only be effective for 
 $w \le 0$.

Eqs.~(\ref {eq:ein1}) and (\ref {eq:ein2}) can now be written as 
\be\label{ees}
H^2=\frac13{\dot\phi}^2+\frac23V\,,\;\;\;\;\;\;\dot{H}=-\dot\phi^2\,,
\ee
for the standard scalar field, and
\be\label{eet}
H^2=\frac23 \frac{U}{\sqrt{1-{\dot{\psi}}^{2}}}\,,\;\;\;\;\;\;\dot H=-\frac{{\dot{\psi}}^{2}}{\sqrt{1-{\dot{\psi}}^{2}}}\,U\,,
\ee
for the tachyon field.

Within the first-order formalism introduced in ref. \cite{bglm}, the correspondence may be obtained explicitly. The starting point is to assume $H$ is a function of either $\phi$ or $\psi$ alone,
or equivalently 
that there is only one value of $H$ corresponding to each value of $\phi$ (or $\psi$).  This assumption is satisfied in many situations of cosmological interest but is not valid in general (for example if $\phi$ (or $\psi$) is oscillating around a minimum of the potential). However, it will be satisfied in the case of slow-rolling scalar fields.
Using Eqs.~ (\ref{ees}) and (\ref{eet})  one obtains
\bes
\ben
\label{dphi}
\dot{\phi}=-H_{,\phi}\,,
\\
\label{dpsi}
\dot{\psi}=-\frac23\frac{H_{,\psi}}{H^2}\,,
\een
\ees
and the potentials
\bes
\ben
\label{v}
V(\phi)=\frac32H^2-\frac{(H_{,\phi})^2}{2}\,,
\\
\label{u}
U(\psi)=\frac32 H^2\sqrt{1-\frac49\frac{(H_{,\psi})^2}{H^4}}\,,
\een
\ees
for the standard and tachyon scalar fields, respectively. 

The correspondence between $\phi$ and $\psi$ can be made explicitly by taking into account that Eqs.~(\ref {dphi}) and (\ref {dpsi}) imply that
\be
\label{dphidpsi}
\frac{d\phi}{d\psi}= \pm \sqrt{\frac32} H\,,
\ee
so that
\bes
\ben
\label{cst}
\phi=\pm  \sqrt{\frac32} \int H d\psi\,,
\\
\psi=\pm  \sqrt{\frac23} \int \frac{d\phi}{H}\,.
\een
\ees

Expanding Eq.~ (\ref{u}) up to first order terms in  $(H_{,\psi})^2/H^4$ and using 
Eq.~ (\ref{dphidpsi}) one obtains
\be
U=  \frac32 H^2-\frac29\frac{(H_{,\psi})^2}{H^2}=\frac32H^2-\frac{(H_{,\phi})^2}{2}=V\,.
\ee
Consequently, for slow-rolling fields, the scalar field potentials $V$ and $U$ are approximately the same.

Note that the knowledge of $H(\phi)$ ( or $H(\psi)$) completely determines $V(\phi)$ ( or $U(\psi)$) but the reverse is not true. This is a result of the freedom to fix the kinetic energy of the scalar field at a given initial time, for a given scalar field potential. Given $V(\phi)$ and a particular $H(\phi)$, the solution ${\bar H}(\phi)=H(\phi)+\Delta H(\phi)$ is also possible with the same potential as long as
\be
\frac32 (\Delta H)^2 + 3 H \Delta H - \frac12 ({\Delta H}_{,\phi})^2 +  H_{,\phi} {\Delta H}_{,\phi}=0\,.
\ee
However, in general, this change in $H(\phi)$ will lead to a modification of the corresponding potential.

\section{Dynamics of inflation}

Up to this point the discussion has been fairly general with few assumptions being made about the dynamics of the universe. In this section we shall investigate the correspondence between standard and tachyon scalar field cosmologies during an inflationary phase in the early universe.  This is essential for establishing a correspondence between the two theories since the classical fluctuations we observe in the universe today are expected to have been generated during inflation.

\subsection{Slow-roll parameters and sound speed}

We start by defining a set of first order slow-roll parameters, expressed as a function of $H$ and its time derivatives, as \cite{wcw}
\begin{equation}
\epsilon_{n}\equiv \frac{\left(-1\right)^{n}}{H}\frac{d^{n}H}{dt^{n}}\times \left(\frac{d^{n-1}H}{dt^{n-1}}\right)^{-1}\,,\label{eq:definition}\end{equation}
where $n \ge 1$and $d^0H/dt^0=H$.
These parameters can be written either as a function of the standard
or the tachyonic scalar field, namely 
\begin{equation}
\epsilon_{1}=-\frac{\dot{H}}{H^{2}}=\frac{H_{,\phi}^{2}}{H^{2}}=\frac{2}{3}\frac{H_{,\psi}^{2}}{H^{4}}\,\label{eq:e1}
\end{equation}
\begin{equation}
\epsilon_{2}=\frac{\ddot{H}}{H\dot{H}}=-\frac{2H_{,\phi\phi}}{H}=-\frac{4}{3}\left[\frac{H_{,\psi\psi}}{H^{3}}-\frac{H_{,\psi}^{2}}{H^{4}}\right]\,,\label{eq:e2}
\end{equation}
\begin{equation}
\epsilon_{3}=-\frac{\overset{...}{H}}{H\ddot{H}}=\frac{2H_{,\phi\phi}}{H}+\frac{H_{,\phi}H_{,\phi\phi\phi}}{HH_{,\phi\phi}}=\frac{4H^{2}H_{,\psi\psi}^{2}+2H^{2}H_{,\psi}H_{,\psi\psi\psi}-16HH_{,\psi}^{2}H_{,\psi\psi}-10H_{,\psi}^{4}}{3H^{4}\left[HH_{,\psi\psi}-H_{,\psi}^{2}\right]}\,.\label{eq:e3}
\end{equation}
The sound speed can be expressed in terms of the slow-roll parameter $\epsilon_1$. Using Eqs. \eqref{eq:new_identifications}, \eqref{lagran}, \eqref{eq:ein2} and  \eqref{eq:e1}  one can show that
\be
{\dot H}=-H^2 \epsilon_1=-2X{\mathcal L}_{,X} \sim  -2 X f\,,
\ee
where the approximation is valid up to first order in $X$. Then, using Eq.\eqref{cs}, one obtains
\be\label{cse}
c_s^2 \sim 1- \frac{2g H^2}{f ^2}  \epsilon_1 \sim 1- c_1 \epsilon_1\,,
\ee
again up to first order in $X$. Here, $c_1$ is a constant which is equal to $0$ for a standard scalar field and $2/3$ for a tachyon field.

\subsection{Scalar and tensor perturbations}

The spectra of scalar and tensor fluctuations at horizon crossing (${\cal P}_{\cal R}$ and ${\cal P}_g$, respectively) produced during inflation has been calculated in \cite{sv} (see also \cite{ll}). Up to first-order in the slow-roll parameter $\epsilon_1$ 
\ben\label{Pr}
{\cal P}_{\cal R}(k)&=&\big[1+2(\alpha+1-c_1)\;\epsilon_1+\alpha\; \epsilon_2 \big]\times\frac{H^2}{8\pi^2\epsilon_1}\Bigg|_{k=aH}\,,\\
\label{Pg}
{\cal P}_g(k)&=&\left[1-2(\alpha+1)\epsilon_1\right]\times\frac{2H^2}{\pi^2}\Bigg|_{k=aH}\,,
\een
where $\alpha=2-\ln2-\gamma\simeq-0.7296$ and $\gamma$ is the Euler constant . 

The scalar spectral index $n$, the tensor spectral index $n_T$, and the tensor-scalar ratio $r$, defined by
\be
{n}\equiv1+\frac{d\ln{\cal P}_s(k)}{d\ln k}\,,\;\;
{n}_{\cal T}\equiv\frac{d\ln{\cal P}_{g}(k)}{d\ln k}\,,\;\;r=\frac{{\cal P}_g}{{\cal P}_{\cal R}}\,,
\ee
were also calculated in  \cite{sv} up to second order in the slow-roll parameters:
\bes
\be
n=1-2\epsilon_1-\epsilon_2-[2\epsilon_1^2+(2\alpha+3-c_1)\epsilon_1\epsilon_2+\alpha\epsilon_2\epsilon_3]\,,
\ee
\be
{n}_{\cal T}=-2\epsilon_1[1+\epsilon_1+(\alpha+1)\epsilon_2]\,,
\ee
\be
r=16\epsilon_1[1+\alpha\epsilon_2-c_1\epsilon_1]\,.
\ee
\ees
This shows that, up to first-order in the slow-roll parameters, the spectral indexes are not sensitive to sound speed. The difference between two dual theories (at the background level), (a) and (b), only appears at second order in the slow-roll parameters:   ${n}^{a}-{n}^{b}=(c_1^a-c_1^b)\;\epsilon_1\epsilon_2$, ${n}^{a}_{\cal T}-{n}^{b}_{\cal T}=0$, ${r}^{a}-{r}^{b}=16(c_1^b-c_1^a)\;\epsilon_1^2$. If (a) and (b) represent standard and tachyon scalar field inflationary cosmologies then $c_1^a=0$ and $c_1^b=2/3$ which implies that ${n}^{a}-{n}^{b}=-2\;\epsilon_1\epsilon_2/3$, ${n}^{a}_{\cal T}-{n}^{b}_{\cal T}=0$, ${r}^{a}-{r}^{b}=32\;\epsilon_1^2/3$.

\section{Examples}

In this section we shall compute the correspondence, at the background level, between tachyon and standard scalar field cosmologies for two specific classes of models of cosmological interest for the dynamics of inflation.

\subsection{Example 1}

Consider $H(\phi)=A e^{-B\phi}+C$, where $A >0$, $B>0$ and $C \ge 0$ are constants. Then, Eq. (\ref{v}) implies that the 
standard scalar field potential is given by
\be\label{vexp}
V(\phi)=\frac12A^2(3-B^2)e^{-2B\phi}+3ACe^{-B\phi}+\frac32C^2\,,
\ee
with
\be
\phi=\frac1{B}\;\ln\left(AB^2\,(t-t_0) + e^{B \phi_0}\right)\,,
\ee
which simplifies to $\phi= B^{-1} \ln(AB^2\,t)$ if we choose, without loss of generality, initial conditions such that $\phi \to -\infty$ when $t \to 0$. In this case, we have
\be
H=\frac{1}{B^{2}t}+C\,,
\ee 
and we obtain the following power law solution for the evolution of the scale factor with 
the time
\be
a(t)=\left(t/t_0\right)^{1/B^2}e^{C(t-t_0)}\,,
\ee
where the scalar factor has been normalized to unity at the time $t_0 > 0$. Hence, the universe is always accelerated (${\ddot a} > 0$) for $t\geq0$ if $C=0$ and $B < 1$ or if $C>0$ and $B\leq1$. If $C>0$ and $B>1$ then inflation only occurs for $t > t_i=(B-1)/CB^2$.

If $C=0$ and $B\neq\sqrt{3}$, the dual tachyon field is 
\be
\psi=\pm\sqrt{\frac23}\frac{1}{AB}\;e^{B\phi},
\ee
and the corresponding potential is given by
\be\label{u1}
U(\psi)=\sqrt{1-\frac23B^2}\frac{1}{B^2\psi^2}\,.
\ee
In this case the potential \eqref{vexp} reduces to the simple exponential form considered in \cite{ckll}, which corresponds
to the tachyon potential \eqref{u1} studied in \cite{lxl}. On the other hand, if $C\neq0$ and $B=\sqrt{3}$, the first term of the 
potential given by Eq. \eqref{vexp} vanishes. In this case the dual tachyon field is given by
\be
\psi=\pm\sqrt{\frac23}\frac1{BC}\;\ln\left(A + C e^{B \phi}\right)\,,
\ee
and, for $A<<1$, the corresponding potential becomes
\be
U(\psi)=\frac32 C^2\left(1+2A e^{-D\psi}\right)\,,
\ee
with $D=3/\sqrt{2} C$.

The slow-roll parameters for the first example are
\be
\epsilon_{1}=B^{2}/(1+\frac{C}{A}e^{B\phi})^2\,,
\ee
$\epsilon_{2}=-2\epsilon_{1}$ and $\epsilon_{3}=3\epsilon_{1}$. The beginning and the end of inflation is defined by the 
condition ${\ddot a}=0$ (or alternatively $\epsilon_1=1$). For $C>0$ and $B>1$, inflation starts when $t\geq t_i=(B-1)/CB^2$, which corresponds to $\phi_i=\ln\left(A(B-1)/C\right)/B$. The number of e-foldings since the beginning of inflation is given by
\be
N(\phi)=\int_{t_i}^{t}{H dt}=-\int_{\phi_i}^{\phi}{\frac{H}{H,_\phi} d\phi}=\frac{\phi-\phi_i}{B}+\frac{C\left(e^{B\phi}-e^{B\phi_i}\right)}{AB^2}\,,
\ee
with $t \ge t_i$ and $\phi \ge \phi_i$.

\subsection{Example 2}

Here we consider a model with
\be
\label{ex2}
H(\phi)=A\phi^{2n}+B\,,
\ee
where $A>0$, $B\geq 0$ and $n$ are real constants. We shall assume that the model is valid only for $\phi > 0$. From \eqref{v} we obtain the potential
\be
V(\phi)=\frac32\left(A\phi^{2n}+B\right)^2-2n^2A^2\phi^{4n-2}\,,
\ee
If $B=0$ and $n=1/2$ the potential becomes the simplest polynomial potential often employed in the study of chaotic inflation \cite{bgkz}, and if $B=0$ and $n=-1/2$  it reduces to the Harrison-Zel'dovich potential obtained in Ref.\cite{ckll}. 
The evolution of $\phi$ with physical time is given by
\be\
\phi=\left(4n(n-1)A t +\phi_0^{2(1-n)}\right)^{1/(2(1-n))},
\ee
where we take $t_0=0$ and $\phi_0 > 0$. Using  \eqref{ex2} and normalizing the scale factor at $t=0$ so that $a_0=1$, one 
obtains
\be
a(t)=\exp\left(-\left(4n(n-1)A t +\phi_0^{2(1-n)}\right)^{1/(1-n)}/(4nA)+Bt\right)\,.
\ee
If $n > 1$ then $t \to t_*=\phi_0^{2(1-n)}/(4n(1-n)A) < 0$ implies that $\phi \to \infty$ with $H \to \infty$ and $V \to \infty$. If $n < 0$ then when $t = t_*=\phi_0^{2(1-n)}/(4n(1-n)A) < 0$ we have $\phi = 0$ and  ${\dot \phi}=-H_{,\phi}=0$.  At this point the first order formalism described in Sect. III ceases to be valid.  In both cases inflation may only occur at late times, when $V \to {\rm constant}$. If $0<n<1$ then when $t = t_*=\phi_0^{2(1-n)}/(4n(1-n)A) > 0$ we have $\phi = 0$ and  ${\dot \phi}=-H_{,\phi}=0$, if 
$n > 1/2$, or  ${\dot \phi}=-H_{,\phi}=\infty$,  if $n < 1/2$. 

If $n=1$ then
\be
\phi=\phi_0\;e^{-2At}\,
\ee
and
\be
a(t)=\;\exp\left(\frac{\phi_0^2}{4A}\left(1- e^{-4A\;t}\right)+B\;t\right)\,.
\ee
The case with $n=0$ is trivial, with $\phi=\phi_0$, $H=H_0$ and $V=V_0$ at all times.

The slow-roll parameters are given by
\bes
\be
\epsilon_{1}=\frac{4n^2A^2}{(A\phi+B\phi^{1-2n})^2}\,,
\ee
\be
\epsilon_{2}=\frac{-4n(2n-1)A}{A\phi^2+B\phi^{2(1-n)}}\,,
\ee
\be
\epsilon_{3}=\frac{4n(3n-2)A}{A\phi^2+B\phi^{2(1-n)}}\,,
\ee
\ees
and a necessary and sufficient condition for inflation to occur is $\epsilon_1 < 1$ or equivalently $A\phi+B \phi^{1-2n} < 2 |n| A$.

The relation for the dual tachyon field, in terms of the standard scalar field, depends of $n$ and B.
For $n\neq1/2$ and $B=0$, the dual tachyon field is
\be
\psi=\pm\sqrt{\frac23}\frac{\phi^{1-2n}}{A(1-2n)}\,,
\ee
and the corresponding potential is
\be
U(\psi)=\frac32A^2(D\psi)^{4n/(1-2n)}\left(1-\frac83n^2(D\psi)^{2/(2n-1)}\right)^{1/2}\,,
\ee
where $D=\sqrt{3/2}\;A(1-2n)$, which, recently \cite{cht} was considered to describe intermediate inflation, for $0<1/(1-n)<1$.
For $n=1/2$, the dual tachyon field is
\be
\psi=\pm\sqrt{\frac23}\frac1A\ln(A\phi+B)\,,
\ee
and the corresponding potential is
\be
U(\psi)=\frac32 e^{2D\psi}\left(1-\frac23A^2e^{-2D\psi} \right)^{1/2}\,,
\ee
where $D=\pm\sqrt{3/2}\;A$, which for $D>0$ reduces to the exponential form at large $\psi$.
For $n=1$ and $B\neq0$, the dual tachyon field is
\be
\psi=\pm\sqrt{\frac2{3AB}}\arctan\left(\sqrt{\frac{A}{B}}\;\phi\right)
\ee
and, for $|B|\geq2/3$, the corresponding potential is
\be
U(\psi)=\frac32B^2\sec^4(D\psi)\left( 1-\frac23\frac{A}{B}\sin^2(2D\psi) \right)^{1/2}\,,
\ee
where $D=\sqrt{{3AB}/{2}}$.

\section{Ending Comments} 

In this paper we have shown that, as long as $w \le 0$, for each tachyon scalar field cosmology there is a standard scalar field model which leads to the same background evolution, and we have explicitly determined the associated transformation between the models. This correspondence, is broken at a perturbative level for $w \neq -1$ since the models have different sound speeds ($c_s^2=1$ and $c_s^2=-w$ for standard and tachyon models, respectively). Still, we have demonstrated that for $w \sim -1$, the correspondence remains valid, even at a perturbative level, up to first order in the slow-roll parameters. In particular, in the inflationary regime the two models generate identical spectra of scalar and tensor perturbations, up to first order in the slow-roll parameters, thus leading to very similar cosmological signatures. We have explicitly determined the correspondence, at the background level, between tachyon and standard scalar field cosmologies for two specific examples of cosmological interest.
\\
The authors would like to thank CAPES, CNPq, FAPESP, Brasil and FCT, Portugal for partial support.


\end{document}